\documentclass[%
 aip,
 amsmath,amssymb,
 reprint,%
longbibliography
]{revtex4-1}

\usepackage{graphicx, import}% Include figure files
\usepackage{dcolumn}% Align table columns on decimal point
\usepackage{bm}% bold math
\usepackage{color}
\usepackage{soul}

\usepackage[utf8]{inputenc}
\usepackage[T1]{fontenc}

\usepackage{tikz}
\usepackage{xcolor}
\definecolor{lightgrey}{rgb}{0.88,0.88,0.88}
\definecolor{lightred}{rgb}{1,0.85,0.88}

\newcommand{\pos}[1]{{\bm #1}}

\begin{document}

%\preprint{APS/123-QED}

\title{Adversarial Reverse Mapping of Equilibrated Condensed-Phase Molecular Structures}

% Shortitle
% \title{Adversarial Backmapping of Molecular Structures}
% Force line breaks with \\

%\thanks{A footnote to the article title}%

\author{Marc Stieffenhofer}
 %\altaffiliation[Also at ]{Max Planck Institute for Polymer Research, Ackermannweg 10, 55128 Mainz}%Lines break automatically or can be forced with \\
%\author{Second Author}%
% \email{Second.Author@institution.edu}
\affiliation{%
 Max Planck Institute for Polymer Research, 55128 Mainz, Germany
}%

%\collaboration{MUSO Collaboration}%\noaffiliation

\author{Michael Wand}
 %\homepage{http://www.Second.institution.edu/~Charlie.Author}
\affiliation{
    Institute of Informatics, Johannes Gutenberg University, 55099 Mainz, Germany
}%

\author{Tristan Bereau}
\affiliation{%
 Max Planck Institute for Polymer Research, 55128 Mainz, Germany
}%
\affiliation{
    Van 't Hoff Institute for Molecular Sciences and
    Informatics Institute, University of Amsterdam, Amsterdam 1098 XH, The
    Netherlands
}

%\collaboration{CLEO Collaboration}%\noaffiliation

\date{\today}% It is always \today, today,
             %  but any date may be explicitly specified

\begin{abstract}
A tight and consistent link between resolutions is crucial to further
expand the impact of multiscale modeling for complex materials. We
herein tackle the generation of condensed molecular structures as a
refinement---backmapping---of a coarse-grained structure. Traditional
schemes start from a rough coarse-to-fine mapping and perform further
energy minimization and molecular dynamics simulations to equilibrate
the system. In this study we introduce DeepBackmap: A deep neural
network based approach to directly predict equilibrated molecular
structures for condensed-phase systems. We use generative adversarial
networks to learn the Boltzmann distribution from training data and
realize reverse mapping by using the coarse-grained structure as a
conditional input. We apply our method to a challenging
condensed-phase polymeric system. We observe that the model trained in
a melt has remarkable transferability to the crystalline phase. The
combination of data-driven and physics-based aspects of our
architecture help reach temperature transferability with only limited
training data.
\end{abstract}

% \pacs{Valid PACS appear here}% PACS, the Physics and Astronomy
                             % Classification Scheme. 
%\keywords{Suggested keywords}%Use showkeys class option if keyword
                              %display desired
\maketitle

%\tableofcontents

\section{Introduction}
\label{sec:level0}

Computational modeling of soft-matter systems inherently requires the
consideration of a wide range of time and length scales, where
microscopic interactions can impact meso- to macroscopic changes.
\cite{peter2009multiscale} Setting aside quantum mechanics, even
molecular dynamics (MD) quickly reaches its limits when probing long
relaxation times. Circumventing such limitations remains an area of
active research, motivated in part by the promises of computational
soft materials discovery.\cite{bereau2016research}  Various strategies
aim at breaking the natural limitations of MD, from enhanced-sampling
techniques \cite{mitsutake2013enhanced} to
dedicated hardware~\cite{shaw2009millisecond} to hierarchical
multiscale modeling.\cite{kremer2002multiscale,
horstemeyer2009multiscale, peter2009multiscale}

Multiscale modeling relies on several levels of resolution, striving
to make best use of each level. At the lower end, a coarse-grained
(CG) resolution will map groups of atoms to single interaction sites
or beads. The CG model aims at reproducing specific features of the
high-resolution model, such as structure or
thermodynamics.\cite{voth2008coarse, noid2013perspective,
brini2013systematic} The reduced representation eliminates some
molecular friction, smoothens the energy landscape, and thereby
effectively accelerates sampling of the conformational space.

While mapping from fine to coarse is straightforward, going the
reverse way is no trivial task. Backmapping means reintroducing lost
degrees of freedom: from CG beads to atoms. The reduced CG resolution
implies that one CG configuration will correspond to an
\emph{ensemble} of atomistic microstates. Ideally, the CG model should
perfectly reproduce the Boltzmann distribution of the atomistic system
along the CG degrees of freedom---the many-body potential of mean
force. As such, backmapping aims at generating an atomistic structure
drawn from the probability distribution of atomistic microstates,
given the CG configuration.

\begin{figure}[htbp]
    \center
    \includegraphics[width=0.9\linewidth]{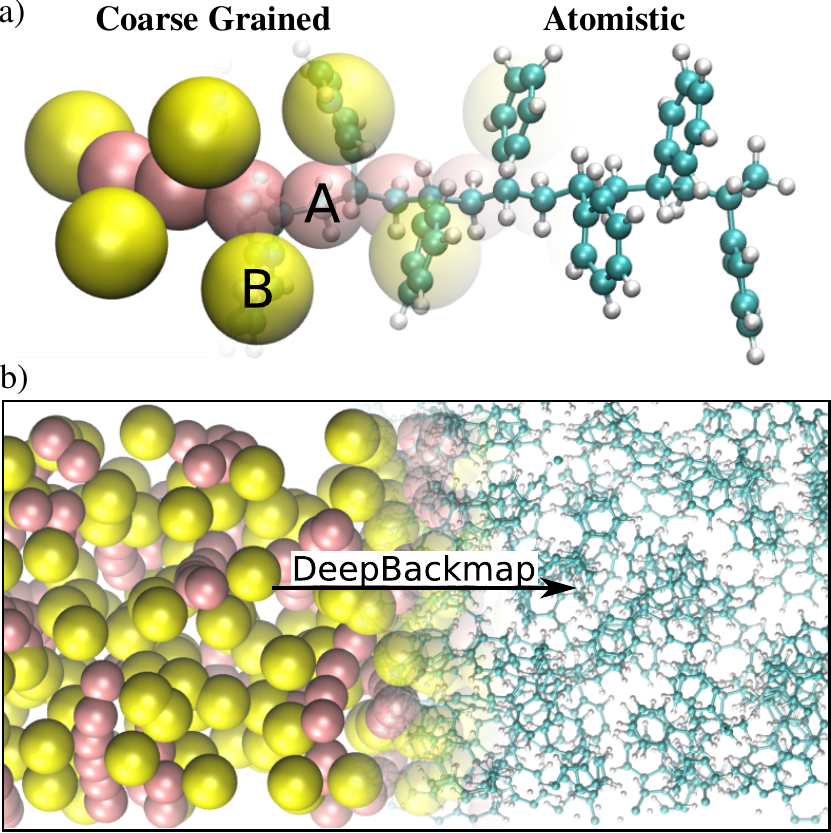}
    \caption{(a) Backmapping consists of reintroducing missing degrees of freedom from a coarse-grained to an atomistic resolution. (b)  DeepBackmap generates Boltzmann-equilibrated atomistic structures conditional on the coarse-grained configuration using an adversarial network. We apply it to the backmapping of a condensed-phase molecular system made of polystyrene chains.}
    \label{intro} 
\end{figure}

The general strategy of existing backmapping schemes is to insert an
initial set of atomistic coordinates into the coarse-grained
structure.\cite{tschop1998simulation2} Two major approaches are random placement of the
atoms close to their corresponding coarse-grained bead center
\cite{rzepiela2010reconstruction,wassenaar} or inserting presampled fragments from a correctly
sampled distribution of all-atom structures.\cite{hess2006long,peter2009multiscale,
zhang2019hierarchical} In both cases energy minimization is required
to relax the initial atomistic configuration and a subsequent
molecular dynamics simulation has to be performed to equilibrate
the system to obtain the correct Boltzmann distribution. 

The computational cost for the subsequent minimization and
equilibration procedures can become significant for high-dimensional
systems. This is also true for backmapping large numbers of
coarse-grained configurations.\cite{menichetti2018efficient}
Furthermore, generating the initial atomistic structure often requires
human intuition to avoid trapping in local minima. For example, the
protocol of Wassenaar \emph{et al.}~needs to introduce geometric
modifiers to correctly reproduce the distribution of torsion angles in
phospholipids.\cite{wassenaar}

In this work we introduce DeepBackmap, a backmapping scheme based on
deep convolutional neural networks (deep CNNs). We bypass
computationally expensive energy minimization and molecular dynamics
simulations by predicting equilibrated atomistic structures directly
from the coarse-grained configuration. This is achieved using
generative adversarial networks (GANs),\cite{gan,
arjovsky2017wasserstein, NIPS2017_7159} a particular type of
generative model based on deep networks: During training, an auxiliary
critic learns a distance metric between generated and training
data. While the critic is trained to maximize the distance, the
objective of the generator is to minimize it. 

The seemingly unintuitive training protocol of GANs circumvents the hard problem of fitting the posterior distribution to training data. Instead of explicitly learning the distribution, they only tune a sampler (the generator) to produce samples indistinguishable from the training distribution (for the critic).
For high-dimensional data sets, such as the joint distribution of many atoms in molecules, previous methods become either intractable or lose resolution, dependencies, or both.

To extend GANs to a conditional model, an auxiliary input can be introduced to both, the generator and the critic, which is taken to be the conditional variable. \cite{isola2017image, mirza2014conditional}
Here we use conditional GANs to learn a coarse-to-fine mapping that re-introduces degrees of
freedom with the correct statistical weight. To this end, we use the coarse-grained structure as an auxiliary input. 

Generating low-energy geometries for molecular compounds remains a
challenge that is still tackled largely by MD simulations. Recent
approaches using machine learning (ML) include autoregressive models,
\cite{Oord16,gebauer2018generating} invertible
neural network,\cite{noe2018boltzmann} Euclidean distance matrices,\cite{hoffmann2019generating} and graph neural networks.\cite{mansimov2019molecular}

Our study uses a convolutional GAN, which has shown the ability to model highly complex and detailed probability distributions (statistical dependency structures) in computer vision applications.\cite{pggan} However, it requires a regular discretization of 3D space, prohibiting scaling to larger spatial structures. We therefore combine the convolutional generator with an autoregressive approach that, in an outer loop, reconstructs the fine-grained structure incrementally, atom by atom. In each step, we use only local information, making the method scalable to arbitrary system sizes. Our method can be used to generate near-equilibrium configurations for condensed-phase systems. 

Backmapping molecules in vacuum can be relatively straightforward, but
the challenge is to achieve it in a condensed phase. We test our
approach on a dense polymeric system: syndiotactic polystyrene (sPS).
sPS not only shows complex structural features in the amorphous melt,
it can also crystallize.\cite{schellenberg2009syndiotactic,
fritz2009coarse, liu2018polymorphism} An illustration of the
coarse-grained and the atomistic representation of the molecule can be
found in Fig.~\ref{intro}. When trained solely on data obtained from a
high-temperature melt, the model is transferable to lower temperatures
where the system is in a crystalline phase. This indicates that the
microscopic degrees of freedom learned by the model have weak
temperature dependence and can be generated solely from large-scale
features captured in the coarse-grained structure.

%\begin{figure}[h]
%  \centering
%  %\def\svgwidth{\textwidth}\footnotesize
%  \def\svgwidth{300bp}
%  %\input{representation.pdf_tex}
%  \import{pdf_tex/}{representation.pdf_tex} 
%  \caption{\label{representation}  }
%\end{figure}

%\def\svgwidth{100bp}
%\input{representation.pdf_tex}

\section{Machine learning model}
\label{sec:level1}

In the following, we discuss our approach. We start with a description of the molecular simulation scenario our method handles, and then discuss in detail how the deep backmapping algorithm works.

\subsection{Setup}
\begin{figure}[htbp]
    \center
    \includegraphics[width=0.8\linewidth]{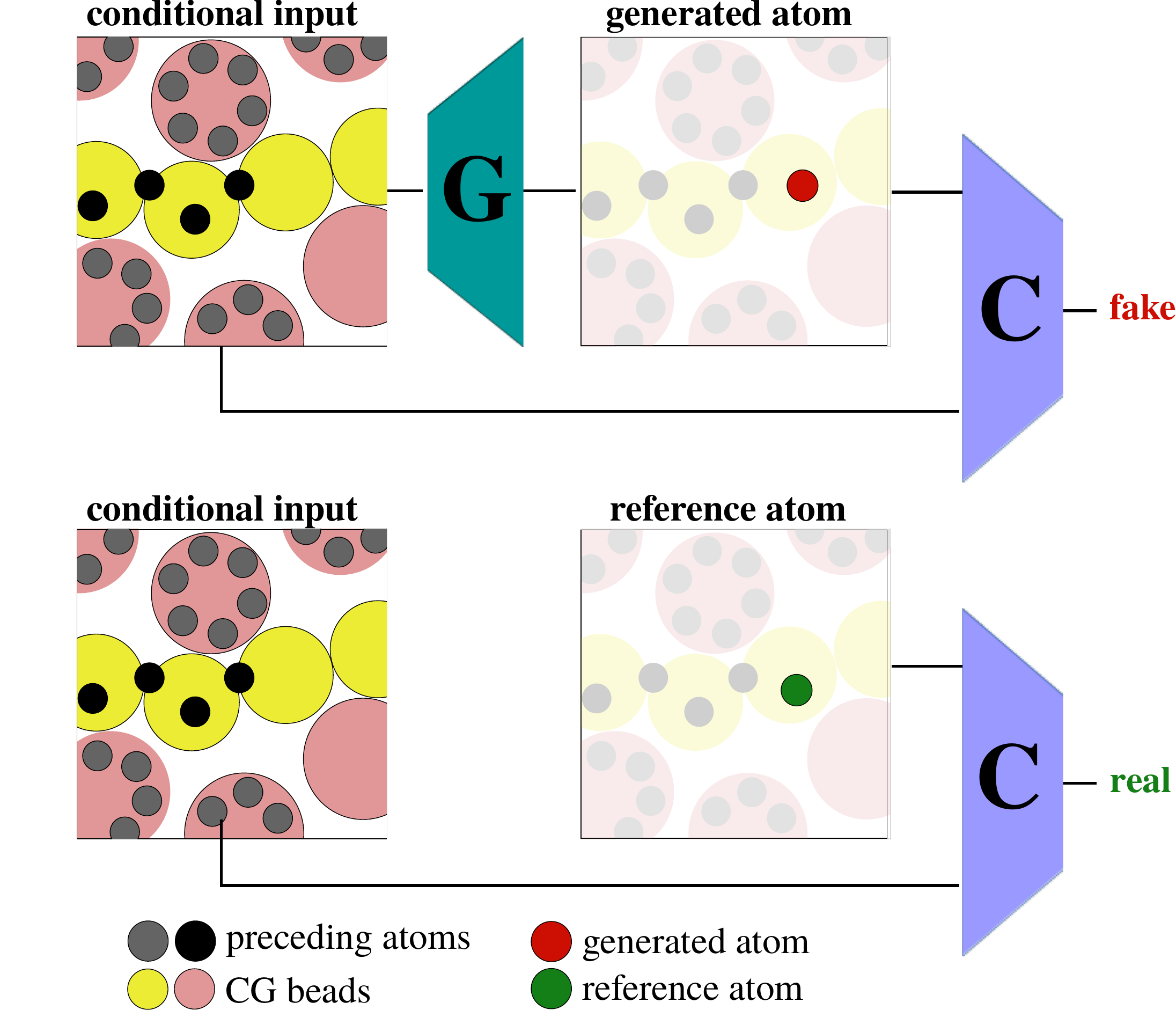}
    \caption{Adversarial autoregressive approach: The generator, $G$,
    sequentially samples atom positions
    conditional on the CG structure and the existing atoms. A critic
    network, $C$, estimates the discrepancy between reference and
    generated atoms.}
    \label{intra_method}
\end{figure}

We define notation for the coarse-grained and atomistic resolutions,
as well as the backmapping procedure:
\begin{description}
\item [Coarse-grained resolution] Let $\{{\bm A}_I = (\pos{R}_I, C_I) \vert
I=1,\dots,N\}$ denote the set of $N$ coarse grained beads. Each bead
has position $\pos{R}_i \in \mathbb{R}^3$ and bead type $C_i$.
\item [Atomistic resolution] Let $\{{\bm a}_i = (\pos{r}_i, c_i) \vert
i=1,\dots,n\}$ denote the set of $n$ atoms, with position $\pos{r}_i
\in \mathbb{R}^3$ and atom type $c_i$. We denote $\varphi_I \subset
\{{\bm a}_i | i=1,\dots,n\}$ as the set of atoms contained in the
coarse-grained bead ${\bm A}_I$.
\item [Backmapping] Backmapping requires us to generate a set of $n$
atom positions $\pos{r}_1,\dots,\pos{r}_n$ conditional on the
coarse-grained (CG) structure, given by the $N$ beads $A_1, \dots ,
A_N$, as well as the atom types $c_1, \dots, c_n$. We express this
problem as a conditional probability $p(\pos{r}_1,\dots,\pos{r}_n |
c_1, \dots, c_n, {\bm A}_1, \dots,{\bm A}_N)$. 
\end{description}
We now propose a machine learning (ML) technique that takes examples
of corresponding coarse- and fine-grained examples as input and from
this \textit{training data} learns the conditional distribution $p$.
Specifically, we do not learn $p$ directly, which is well-known to be
a hard problem for high-dimensional phase spaces,\cite{gan} but rather
infer a sampler that can generate further samples from $p$, see
Fig.~\ref{intra_method}. 

\subsection{Outer loop: Autoregressive model}

Rather than learning to sample from  $p(\pos{r}_1,\dots,\pos{r}_n
\vert c_1, \dots , c_n, {\bm A}_1, \dots, {\bm A}_N)$ directly, we
propose to factorize $p$ in terms of atomic contributions, where the
generation of one specific atom becomes conditional on both CG beads
as well as all the atoms previously
reconstructed.\cite{gebauer2018generating} Based on this factorization
we can train a generative network, $G$, to generate and refine the
atom positions sequentially.

The backmapping scheme hereby consists of two steps: ($i$) An initial
structure is generated using the factorization
\begin{widetext}
\begin{equation}
    \label{factorization}
    p(\pos{r}_1,\dots,\pos{r}_n \vert c_1, \dots , c_n, 
    {\bm A}_1, \dots, {\bm A}_N) 
    = \prod_{i = 1}^n p\left({\pos{r}}_{S(i)} \vert
    \pos{r}_{S(1)},\dots,\pos{r}_{S(i-1)},
    c_{S(1)},\dots,c_{S(i)}, 
    {\bm A}_1,\dots, {\bm A}_N\right), 
\end{equation}
\end{widetext}
where $S$ sorts the atoms in the order of reconstruction and
$\{\pos{r}_{S(1)},\dots,\pos{r}_{S(i-1)}\}$ correspond to atoms that
have been already reconstructed. The dependence on earlier predictions
of $G$ makes our approach \emph{autoregressive}. This procedure would
be exact in a Markovian regime where each atom interacts directly only with its predecessor and successor
%provided statistical independence ofthe atoms 
(so-called ``chain structures''\cite{Koller2009}).
Unfortunately the complexity of condensed-phase liquids calls for more
feedback to avoid steric clashes; $(ii)$ Intuitively, we cannot
optimally place an atom without its whole environment present. This
issue is compounded for ring-like structures, like the phenyl group in
polystyrene. To this end we perform a variant of Gibbs sampling, which
iteratively resamples along the sequence $S$ several
times.\cite{GemanGeman84} Each further iteration still updates one
atom at a time, but uses the knowledge of \emph{all} other atoms.
Experiments confirmed that such Gibbs sampling leads to a good
approximation of $p$, even with a small number of iterations and
fixing the atom ordering.

%To  this end we perform Gibbs sampling to refine the atom positions and update the atom positions according to $p\left({\bm r}_{i} 
% | \{ {\bm r}_{j} \}_{j \neq i}, \{ {\bm R}_I \}_{I=1}^{N}\right)$.

%Because ML models work more efficiently with
%smooth input data, we will represent atoms and CG beads with a smooth
%density, $\gamma({\bm x})$ and $\Gamma({\bm x})$, respectively. This
%amounts to learning the proxy probability distribution
%$p(\gamma|\Gamma)$.

%Based on this factorization we can train a generative network, $G$, to
%generate and refine $\{ {\bm \hat{r}}_i \}_{i=1}^{n}$ atom by atom.
%\begin{equation}
%    G: (\{\gamma_{S(j)} \}_{j=1}^{i-1} ,\Gamma) \mapsto \hat\gamma_i.
%\end{equation}

\subsection{Representation}

Iterative sampling algorithms, such as the Gibbs sampler, have high
computational cost. We hereby optimize our approach by means of a
robust learning algorithm that can capture complex dependencies in the
local environment directly.

The problem of learning complex, high-dimensional and high-order
dependencies in generative models has received considerable attention in computer vision.
The most successful technique for this task are generative deep
convolution neural networks\cite{fukushima1980neocognitron} (deep
CNNs) trained by adversarial training.\cite{gan,pggan} There is also
growing evidence that deep networks are also effective in capturing
the statistics of physical
systems.\cite{noe2018boltzmann,gebauer2018generating}
%\cite{mccann2017convolutional, rawat2017deep}% \tb{What is a physical
ensemble?
%Rephrase}.

%Generating a 3D structure requires a representation that captures
%spatial information and encodes the mutual orientation of the
%particles. Inspired by the great success of convolutional neural
%networks (CNNs) for image-generation tasks \cite{mccann2017convolutional, rawat2017deep} \tb{Refs}, 

In order to leverage deep CNNs for our task, an explicit spatial discretization of the ambient space, similar to pixels in an image, is required. The standard technique is to use a voxel-based representation.\cite{Wu2015} To this end, we represent atoms and CG beads with a smooth
density, $\gamma({\bm x})$ and $\Gamma({\bm x})$, respectively.

The particle densities are modeled using Gaussian distributions, such that for atom $i$ we define
\begin{equation}
    \gamma_i ({\bm x}) = \exp \left( - \frac{({\bm x} - {\bm r}_i)^2}{2\sigma ^2} \right),
\end{equation}
where ${\bm x}$ is the spatial location in Cartesian coordinates,
expressed on a discretized grid due to the voxel representation. The
density is centered around particle position ${\bm r}_i$ with Gaussian
width $\sigma$, treated as a hyper parameter. CG beads are similarly represented.

\subsubsection{Locality}

The high costs of large regular 3D grids are the reason for employing deep CNNs only locally and using the previously described outer loop to build-up larger structures incrementally using autoregressive sampling.
To make the model scalable to large system sizes, we assume locality by limiting the information about the
environment to a cutoff $r_\textup{cut}$. 

We encode the local environment of an atom $i$ or CG bead $I$ by means
of the density of particles placed around it, denoted $\xi_{i,I}$ and
$\Xi_{I}$, respectively. We sum over all atoms or beads within a cubic
environment of size $2 r_{\mathrm{cut}}$. We shift all atom and bead
positions around the CG bead of interest, $I$. Further, we rotate the
local environment to a local axis system. This improves generalization
from limited training examples by removing three translational and
two of the rotational degrees of freedom, i.e., the ML algorithm does 
not need to learn the corresponding coordinate invariance from (additional) examples.

Specifically, we align the bond between consecutive CG beads $I-1$ and
$I$ to the local $z$ axis using a rotation matrix $M_I$ to construct
the local environment of atom $i$
\begin{equation}
 \xi_{i,I}({\bm x}) = \sum_{j=0}^{i-1} \gamma_{S(j)} 
    (M_I({\bm x} - {\bm R}_I)),
\end{equation}
which extends over the region $-r_\textup{cut} < x_\alpha <
r_\textup{cut}$ and $\alpha$ runs over the three Cartesian
coordinates. Similarly the coarse-grained environment is constructed
as
\begin{equation}
 \Xi_{I}({\bm x}) = \sum_{J=0}^{N} \Gamma_J
    (M_I({\bm x} - {\bm R}_I)).
\end{equation}
In this work we set $r_\textup{cut} = 6$\,\AA, such that several CG
beads are included in the local environment (see Fig.~S2).
Importantly, $\xi_i$ and $\Xi_I$ are discretized on a regular grid.

\subsubsection{Feature embedding}
%\ms{please read and see if this is ok}
%\mw{I find the notion of the ``feature axis'' very confusing. This is a discrete index to enumerate the particles, isn't it? Also, Deep NNs barely use traditional features but rather learn the representation from raw data (``representation learning''). The main point here is to use per-particle densities as input, on which the representation is then learned. tl;dr -- I would reconsider the feature axis thing.}

%ML algorithms require features as input to the trained predictor. In the case of CNNs, most of the feature representation is learned automatically by the multi-layered architecture (``representation learning''). However, we need to feed geometric data (3D images, in our case) into the network as primary information source. 

%A CNN requires training data to be input in the form of ``images'', i.e., vector-valued functions $f:\mathbb{R}^{d_i} \rightarrow \mathbb{R}^{d_o} $ on a low-dimensional domain (i.e., $d_i$ is small, usually $d_i=1..3$; $d_o$ may be large.). In our case, we use a 3D input domain, represented as voxels, and define a set of $d_o$ separate scalar fields to represent the atomic neighborhood of an atom in the input molecule. In other words, every voxel stores a vector in $\mathbb{R}^{d_o}$; in the following, we call its components ``feature channels''.

A CNN takes an image (typically 2D or 3D) as input where every pixel
or voxel is vector-valued. For example, an RGB image consists of three
\textit{feature channels}: One channel for every primary color. Here,
we store a number of feature channels in each voxel that represent the
presence of other atoms or beads of a certain kind. In the most basic
version, we could use a single feature channel to encode all other
atoms. However, this would make it impossible to distinguish their
type and might also lead to clutter. The opposite extreme would be to
assign a separate feature channel to each atom. The downside here is
not only increased memory costs but, more importantly, the loss of
permutation invariance of the atoms.

\begin{figure*}[htbp]
    \begin{center}
    \includegraphics[width=0.8\linewidth]{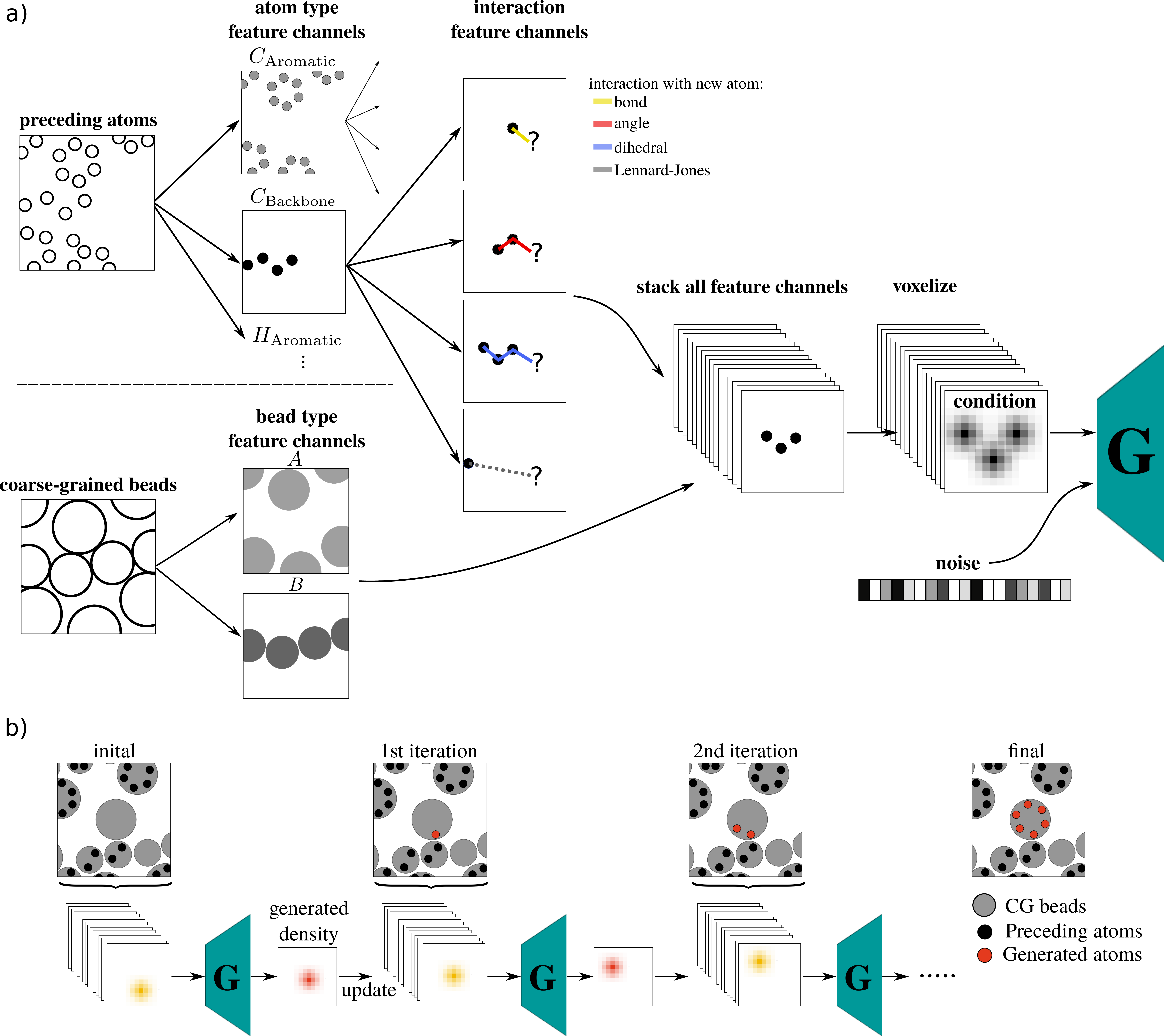}
    \caption{\label{representation} a) Representation and conditional
    input. Existing atoms and CG beads are split into separate
    channels according to their atom/bead type. In addition, the
    atomic information is further split in terms of intra- and
    intermolecular interactions. All channels are used as input for
    the generator network, $G$. b) Recurrent training. Starting from
    an atomistic configuration taken from training data (black) the
    predicted atoms (red) will be added to the local environment
    description for predicting the next atom in the sequence.}
    \end{center}
\end{figure*}

As shown in Figure~\ref{representation}a, we create separate feature
channels for each atom type. Atom types are distinguished not only by
element but additionally by chemical similarity, i.e., atoms of a
given type can be treated as identical in the MD simulation.
Specifically, we classify similarity following the force field for sPS
by Mueller-Plathe.\cite{muller1996local} For atoms of the same type,
we further add channels to distinguish the functional form of
interaction to the current atom of interest. Interaction types
distinguish between bond, bending angle, torsion, and Lennard-Jones.
Similarly, we use separate channels to encode the different
coarse-grained bead types.

Formally, let $f \in \{1,2,\dots,N_F\}$ denote the index of the $N_F$
different feature channels. We define the activation function,
$h_f(S(j))$, to denote association with a channel $f$ 
\begin{equation}
 h_f(S(j)) =
  \begin{cases}
      1,& \text{if atom } S(j) \text{ has feature }f\\
      0,              & \text{otherwise},
  \end{cases}
\end{equation}
and $H_f(J)$ to similarly encode the bead type. We then build a
density map for each channel for both atomic environments
\begin{equation}
 \xi_{i,I}({\bm x},f) = \sum_{j=0}^{i-1} \gamma_{S(j)} 
    (M_I({\bm x} - {\bm R}_I)) h_f(S(j)),
\end{equation}
and coarse-grained environments
\begin{equation}
 \Xi_{I}({\bm x},f) = \sum_{J=0}^{N} \Gamma_J
    (M_I({\bm x} - {\bm R}_I)) H_f(J).
\end{equation}

%\onecolumngrid
%\twocolumngrid

\subsection{\label{sec:level2}Generative model}

Training a generative model is challenging as it requires to measure
and optimize closeness of the target distribution and the
generated distribution of the model. A direct maximum likelihood
training, where the model's parameters are tuned such that the
likelihood of observing the data given the model is optimized, is
infeasible in high dimensions because the normalization factor---the
partition function---cannot be computed efficiently.

Approaches to circumvent these limitations include approximate
techniques like variational autoencoders, where a stochastic lower
bound of the log-likelihood is optimized. Another solution are
likelihood-free methods, such as adversarial training,\cite{gan} that
operate indirectly, by building a sampler and comparing its output to
actual data with a second, ``adversarial'' network. In recent
literature, this approach appears to yield the strongest results, in
particular on high-dimensional and hard to model image
spaces.\cite{pggan} The next best option are auto-regressive models,
which tackle the complexity issue by learning single decisions at a
time.\cite{Oord16} We use this approach in the outer loop but employ
the more expressive GAN for modeling the local placement of atoms. 

Formally, to perform adversarial training, a second network is
introduced, called critic $C$, to distinguish between training samples
and samples from the generative model $G$. The generator competes with
the critic $C$ and is trained to generate samples that $C$ can not
distinguish anymore from training samples. In the conditional
adversarial framework\cite{isola2017image, mirza2014conditional} both
networks $G$ and $C$ are provided with auxiliary information like a
class label to generate samples related to this information. In this
study, we use a conditional generative adversarial network (cGAN) to
generate new atom positions from a random noise vector $z \sim N(0,1)$
and the conditional input $u_i := \{\xi_{i,I}, \Xi_I, c_i \}$
consisting of the local environment representation $\xi_{i,I}$ and
$\Xi_I$, as well as the current atom type $c_i$. In a first step, the generator $G$ predicts a smooth-denisty representation $\hat\gamma_i := G(z,u_i)$.

\subsubsection{From densities to coordinates}

While the smooth-density representation $\hat\gamma_i$
is adequate for a CNN, we ultimately wish to collapse these back to
point coordinates.  We simply compute a weighted average, discretized
over the voxel grid
\begin{equation}
 \hat{\bm r}_i = \int {\rm d}{\bm x}\, \hat\gamma_i({\bm x}) 
 \approx \sum_m \sum_k \sum_l x_{mkl}\hat\gamma_i(x_{mkl}).
\end{equation}
This step is performed for each generated density separately, one atom at a time. We note that
this density-collapse step is differentiable and can thus be easily
incorporated in a loss function.

\subsubsection{Training}

Training of a GAN model is split in two networks: the adversarial critic
and the generative network. The following describes the two loss
functions.

We train a critic network $C$ to distinguish between reference
densities $\gamma_i$ related to the conditional input $u_i =
\{\xi_{i,I}, \Xi_I, c_i \}$ and generated densities $\hat\gamma_i =
G(u_i, z)$.  
The critic aims at both ($i$) distinguishing reference
from generated samples and ($ii$) ensuring smoothness of the
classification with respect to the generator's parameters. 
Both criteria can be fulfilled using a variant of adversarial models where the critic $C$ is used to approximate the Wasserstein distance.\cite{arjovsky2017wasserstein}

The loss function is constructed using the Kantorovich-Rubinstein
duality, which requires $C$ to be constrained to the set of
1-Lipschitz functions. A differentiable function is 1-Lipschitz if and
only if it has gradients everywhere with norm at most one. A soft
version of this constraint is enforced with a penalty on the gradient
norm\cite{NIPS2017_7159}
\begin{align}
    \mathcal{L}_C =\underset{i}{\mathbb{E}}
    \big[  &  C(u_i, \gamma_i) - C\left(u_i, G(u_i, z)\right)\\
    &+ \lambda_\textup{gp} \left(\lVert\nabla _{\tilde{u}_i,\tilde\gamma_i}
    C(\tilde{u}_i,\tilde\gamma_i)\rVert_2-1\right)^2
    \big],
\end{align}
where $(u_i, \tilde{\gamma}_i)$ is interpolated linearly between pairs of points $(u_i, \gamma_i)$ and
$(u_i, G(u_i, z))$. The prefactor $\lambda_\textup{gp}$ scales the weight of the gradient penalty.
%\begin{align*}
%\underset{\Phi}{max} \underset{\mathbf{\hat{y}}}{\mathbb{E}}[D(\mathbf{\hat{y}},\mathbf{x},\mathbf{X})]  -  \underset{\mathbf{y}}{\mathbb{E}}[D(\mathbf{y},\mathbf{x},\mathbf{X})] \\ + \lambda \underset{\mathbf{\tilde{y}}}{\mathbb{E}}[(||\nabla _{\mathbf{\tilde{y}}}D(\mathbf{\tilde{y}})||_2-1)^2]
%\end{align*}
%where $\mathbf{\tilde{y}}$ is sampled uniformly along straight lines between pairs of points $(\mathbf{y}, \mathbf{\hat{y}})$.

\newcommand{\neigborhood}[1]{N_{#1}}

For the generator we combine two aspects to help generate faithful
structures: ($i$) the critic that compares reference and generated
samples, $C\left(u_i, G(u_i, z)\right)$, and ($ii$) a physical prior,
$\Phi$. $\Phi$ aims at accelerating convergence by helping the
generator refine its output. It combines both force-field-based energy
contributions, $E_\textup{FF}$, and a geometric center-of-mass
distance contribution, $d_\textup{COM}$. The prior depends on the set
of atoms corresponding to a coarse-grained bead, $\varphi_I$ for
reference atoms and $\hat{\varphi}_I$ for generated atoms, as well as
reference atoms $N_I$ in the local neighborhood of different beads: 
\begin{equation}
    \Phi(\varphi_I, \hat{\varphi}_I, N_I ) =
    E_\textup{FF}(\varphi_I, \hat{\varphi}_I, N_I) +
    d_\textup{COM}(\varphi_I, \hat{\varphi}_I).
\end{equation}
The force-field-based term penalizes discrepancies between samples
with respect to specific intra- and intermolecular interactions within all neighborhoods $N_I$.
%\tb{$E_\textup{FF}$ depends on atoms beyond one CG bead alone}
\begin{equation}
    E_\textup{FF}(\varphi_I, \hat{\varphi}_I, N_I)
    = \sum_t \left|\varepsilon_{t}(\varphi_I, N_I) - 
    \varepsilon_{t}(\hat{\varphi}_I, N_I)\right|,
\end{equation}
where $t$ runs over the interaction types: intramolecular bond, angle,
and dihedral, and non-bonded Lennard-Jones. The set of
interactions follow the reference atomistic force field.
In the following, let $\theta_I = \{i | a_i \in \varphi_I\}$ be the set of atom indices for atoms contained in $\varphi_I$. The second
term in the physical prior penalizes discrepancies in the
center-of-mass geometry between samples
\begin{equation}
    d_\textup{COM}(\varphi_I, \hat{\varphi}_I)
    = \left| g(\varphi_I) - g(\hat{\varphi}_I) \right|,
\end{equation}
where $g$ refers to the center of mass
\begin{equation}
  g(\varphi_I) = \frac{ \sum_{i\in\theta_I} m_i \pos{r}_i } {\sum_{i\in\theta_I} m_i},
\end{equation}
with $m_i$ being the mass of atom $a_i$.

Overall this leads to the following loss function for the generator
\begin{equation}
    \mathcal{L}_G = \underset{I}{\mathbb{E}}
    \Big[  
        \underset{i \in \theta_I}{\mathbb{E}} 
	    \big[
	      C\left(u_i, G(u_i, z)\right) 
	    \big]
        + \lambda_\Phi 
        \big[
            \Phi(\varphi_I, \hat{\varphi}_I, N_I)
        \big]
    \Big]
\end{equation}
where the prefactor $\lambda_\Phi$ scales the weight of the
physical prior.

The two loss functions, $\mathcal{L}_C$ and $\mathcal{L}_G$ are
trained iteratively and alternatingly until the process reaches equilibrium.
% \ms{nash-equilibrium = convergence?}. % \mw{Does the WGAN still play a game, or is this a convergent objective (i.e., the optimization performs a descent on a potential function rather than following a general vector field)? I am not 100\% sure -- I would therefore leave Mr. Nash out of the discussion here. Addendum: I checked; the story is more complicated -- \url{https://arxiv.org/pdf/1801.04406.pdf}. WGAN-GP is not going down a potential. }

\subsubsection{Implementation details}

We choose a 3D convolutional neural network (CNN) architecture with
residual connections for $ G $ and $ C$.\cite{he2016deep} See Fig.~S6
for a detailed network description. 

The model is trained for 38\,660 iterations in total using a batchsize
of 36. For stability reasons, we start training with $\lambda_\Phi =
0$ and increase it smoothly to $\lambda_\Phi = 0.01$ from step 6000 to
10\,000. Training is performed using the Adam optimizer. The prefactor
scaling the weight of the gradient penalty term is set to
$\lambda_\textup{gp} = 0.1$. To obtain reliable gradients for the
generator, the critic should be trained until optimality. Therefore
the critic $C$ is trained five times in each iteration while the
generator $G$ is trained just once.

We train the model recurrently on atom sequences containing either all
heavy (carbon) or light (hydrogen) atoms corresponding to a single
coarse-grained bead. During training, the initial atomistic
environment representation $\xi_{i,I} $ for each sequence is generated
from training data and contains the atoms present (according to the
order $S$) in the local neighborhood $N_I$ of bead $I$. After each
step, the generated atom density is added to the local environment
representation for the next atom in the sequence, as illustrated in
Fig.~\ref{representation}b, untill all atoms of the sequence are
generated. 

%Note that due to the autoregressive approach the atomistic environment
%representation $\xi_{i,I} $ depends on previous predictions
%$(\hat\gamma_{S(i-1)}, \hat\gamma_{S(i-2)}, \dots)$. We train the model
%recurrently starting from preceding atoms taken from training data.
%After each step the generated atom density $\hat\gamma_{S(i)}$ is
%added to the local environment description for the next atom (see
%figure \ref{representation}b). The sequences we train on are chosen
%such that they contain either all heavy (carbon) or all light
%(hydrogen) atoms corresponding to a single coarse-grained bead
%\tb{Unclear. Please rephrase}. 

In the Gibbs-sampling step, information of all preceding and
subsequent atoms is used to refine the positions of light atoms. On
the other hand for heavy atoms we remove hydrogens from the current
and adjacent beads such that misplaced hydrogens will not hinder $ G$
to find suitable positions for the heavy atoms.

Note that our architecture is not fully rotational equivariant as it
only aligns the region considered by the generator according to the position of the
central bead and the difference vector to the previous bead. This
leaves one rotational degree of freedom around that axis; therefore,
we augment the training set using rotations about that axis. During
prediction we feed different orientations about said axis as well and
choose the structure with the lowest energy from the generated
ensemble.
%Note that the CNN architecture is not rotational equivariant. While rotating the environment to a local axis system fixes the orientation with respect to the x and y axis the environment is still free to rotate about the z-axis. Therefore we augment the training set using rotations about the z-axis. During prediction we feed different orientations about the z-axis as well and choose the structure with the lowest energy from the generated ensemble.

\section{Computational methods}
\label{sec:exp}

\subsection{Reference data}

The atomistic data in this study was reported in Liu \emph{et
al.};\cite{liu2018polymorphism} the underlying force field is based on
the work of Mueller-Plathe.\cite{muller1996local} Replica Exchange MD
simulation, a temperature-based enhanced sampling technique, was used
to sample the system. All simulations were performed using the
molecular dynamics package {\sc GROMACS} 4.6.\cite{hess2008gromacs}
Molecular dynamics simulations are performed in the \emph{NPT}
ensemble using the velocity rescaling thermostat and the
Parrinello-Rahman barostat. An integration timestep of 1~fs is used.
For additional details regarding the simulations the reader is
referred to the work of Liu \emph{et al.}\cite{liu2018polymorphism}
%\ms{that is a bit confusing because Chan used different simulation parameters to sample the system compared to the simulation i did for the equilibration check. She used a berendson barostat and i am not sure about the heat-up in her simulation.} \tb{Then you should rewrite all simulation parameters here. If something's not clear from her paper, email her.}

%Initial velocities are generated according to a Maxwell distribution. Molecular dynamics simulations areperformed in this work were done \emph{without} heat-up
%The total training set consists of twelve different snapshots. 
%To increase the training set and tackle the problem of equivariance with respect to rotations we augment the training set using rotations about the z-axis of the local frames.
%Note that we train the model solely on the high temperature data at $568K$ where the system is in a amorphous phase. Data for different temperatures is used for testing only.

Our training/test data consists of pairs of corresponding fine- and coarse-grained snapshots. To this end, we start from the atomistic frame and apply a fine-to-coarse mapping to obtain the coarse-grained structures.
We use uncorrelated snapshots from three different trajectories
simulated at $T=568$\,K, $453$\,K, and $313$\,K. The system includes
36 polystyrene chains and each chain consists of 10 monomers.

The fine-to-coarse mapping is based on the coarse-grained model
developed by Fritz \emph{et al.}\cite{fritz2009coarse} It represents
the coarse-grained molecule as a linear chain, where each monomer is
mapped onto two CG beads of different types, denoted A for the chain
backbone and B for the phenyl ring (see figure \ref{intro}). Bonds are
created between the backbone beads A-A and between backbone and phenyl
ring beads A-B. The coarse grained model, parameterized in the melt,
is transferable to the crystalline phase and stabilizes the
experimentally observed $\alpha$ and $\beta$
polymorphs.\cite{liu2018polymorphism}

\subsection{Baseline Model}

We compare our results with a generic backmapping scheme developed by
Wassenaar \emph{et al.}\cite{wassenaar} This method places each particle on
the weighted average position of the coarse grained beads it belongs
to and optionally adds a random displacement. The protocol continues
with corrections to the structure using geometric modifiers, setting
the alignment of the next particle as cis, trans, out, or chiral with
respect to the others. Note that those modifiers need first be
manually defined by the user. 

The corrected structure is then relaxed by a force-field based energy
minimization. The first cycle of energy minimization consists of $200$
steps and is performed with non-bonded interactions turned off. The
second cycle of energy minimization consists of $5000$ steps with all
interactions turned on. Clearly the energy minimized structures will
not capture the right Boltzmann distribution and therefore the
protocol of Wassenaar continues with several cycles of position
restrained molecular dynamics simulations. Since we aim for a
backmapping scheme that performs well without running molecular
dynamics simulation, we stop the protocol after the energy minimization
and compare the methods without running any further molecular dynamics
simulations.

\section{Results}

\begin{figure}[htbp]
    \center
    \includegraphics[width=0.80\linewidth]{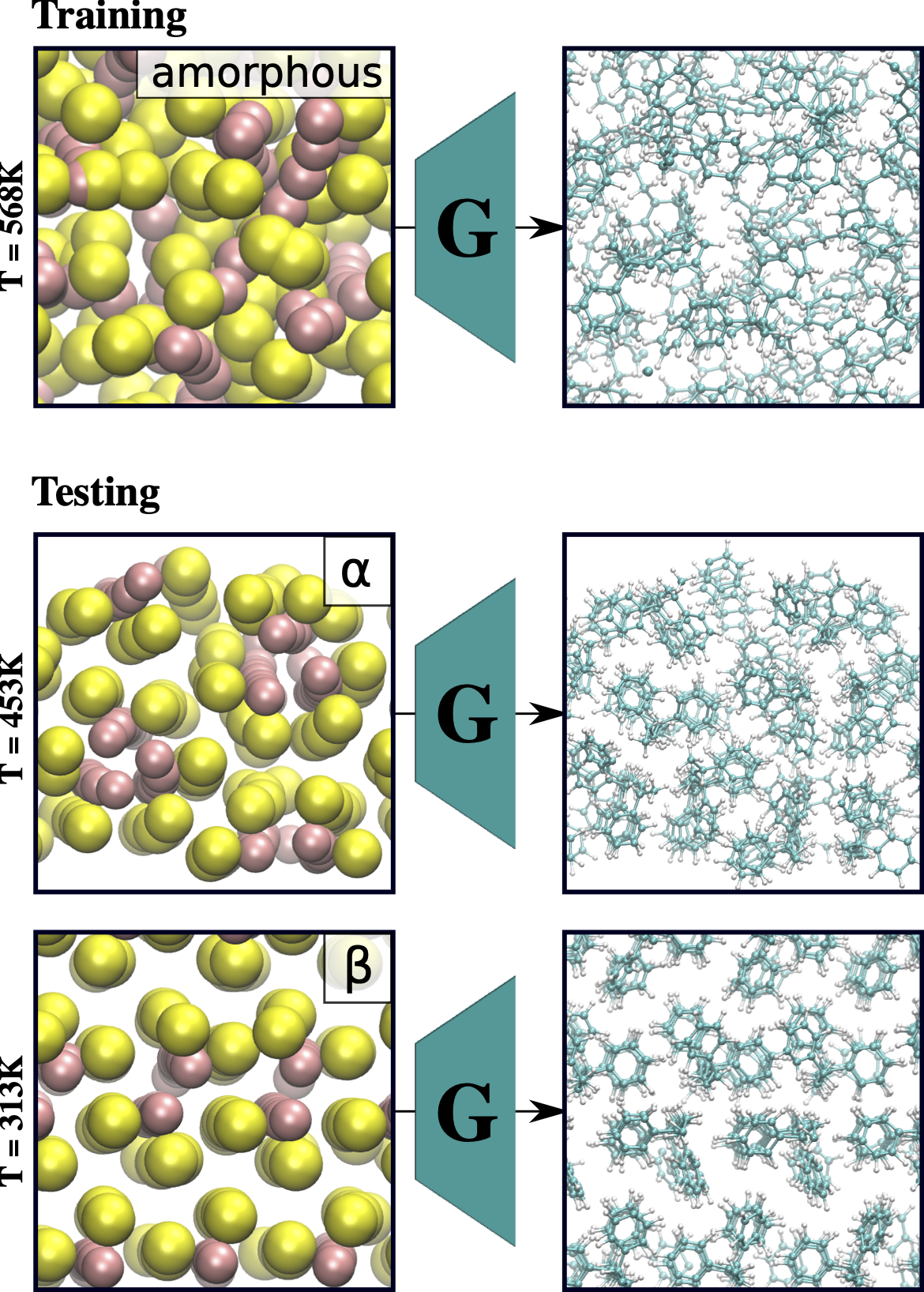}
    \caption{\label{ps_polymorphs} Polymorphism of Polystyrene. At
    high temperature ($T=568$\,K) the system stabilizes an amorphous
    phase. At lower temperatures the CG model mostly stabilizes the
    $\alpha$ polymorph at $T=453$\,K and the $\beta$ polymorph at
    $T=313$\,K. We train DeepBackmap solely on the high-temperature
    ensemble ($T=568$\,K) and test its transferability to the lower
    temperatures. }
 \end{figure}

We apply DeepBackmap to a challenging condensed-phase molecular
system: syndiotactic polystyrene. Despite its simple chemical
structure, polystyrene displays a rich conformational space. Its
syndiotactic form can crystallize, and exhibits complex polymorphic
behavior. Upon thermal annealing, a polystyrene melt undergoes a phase
transition from amorphous to a crystalline phase at $T\approx450$\,K.
The CG model was shown to stabilize the two main crystal polymorphs
$\alpha$ and $\beta$ (see
Fig.~\ref{ps_polymorphs}).\cite{liu2018polymorphism}

We probe the model's ability to transfer across temperatures. To this
end, we train DeepBackmap solely on high-temperature, amorphous
configurations, but validate it at several temperatures
(Fig.~\ref{ps_polymorphs}). The training set consisted of only 12
snapshots simulated at $T = 568$\,K. The model was then applied to MD
configurations at $T=568$\,K, $453$\,K, and $313$\,K, each containing
$78$ samples that were not used during training. For brevity we only
report results about the highest and lowest temperature. We evaluate
the performance of the model regarding its ability to reproduce
structural and energetic features of the reference atomistic
configurations, as well as a comparison with the baseline method.

\subsection{Local structural and energetic features}

\begin{figure}[htbp]
    \begin{center}
    \includegraphics[width=1.0\linewidth]{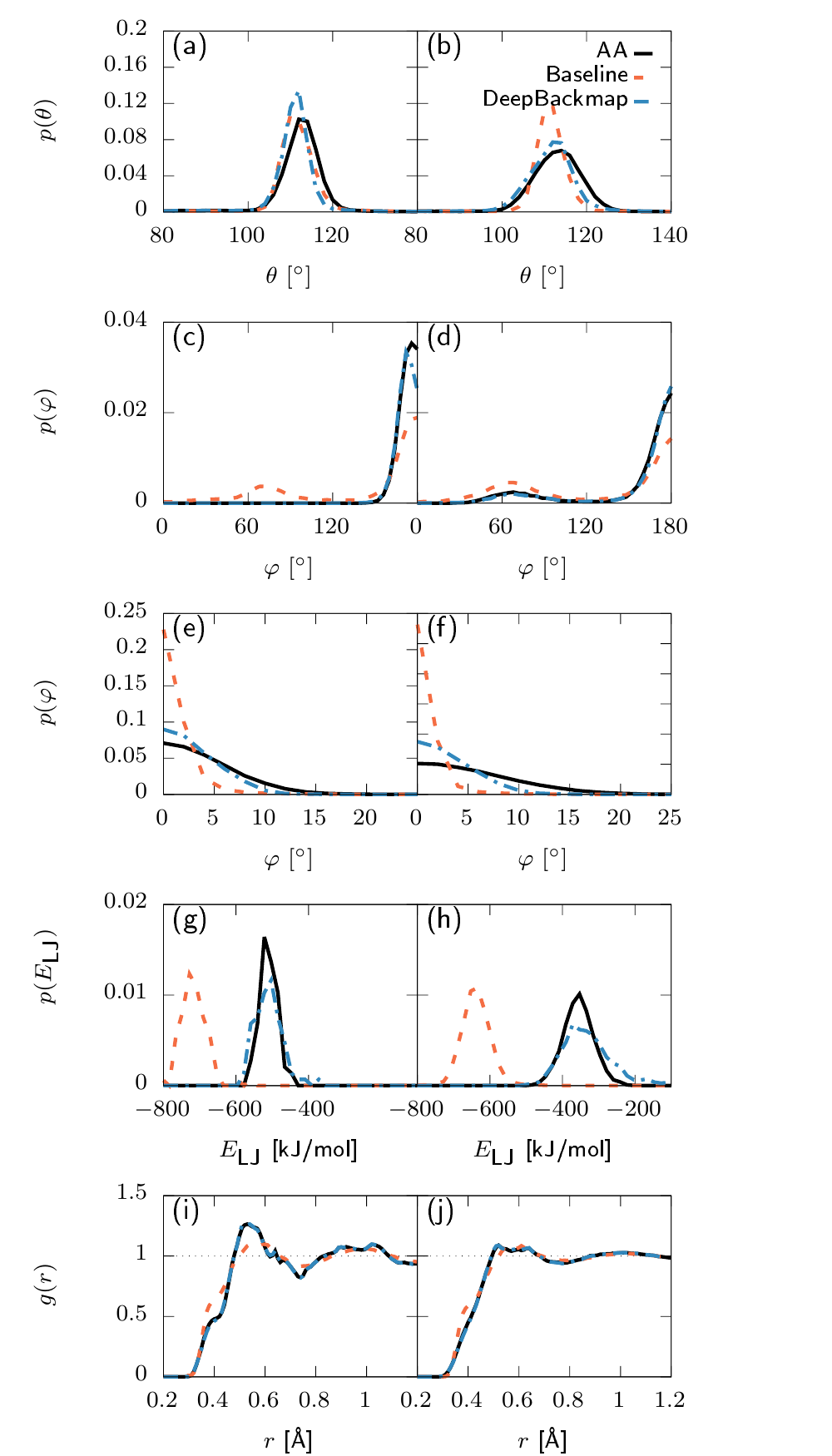}
    \caption{Canonical distributions for various force-field
    interaction terms at (left) $T=313$~K and (right) $T=568$~K. (a-b)
    C-C-C backbone angle, (c-d) C-C-C-C backbone dihedral, (e-f)
    C-C-C-C improper dihedral, (g-h) Lennard-Jones energies, and (i-j)
    radial distribution functions, $g(r)$, of the non-bonded carbon
    atoms. %{\color{blue}TB: The $g(r)$ features are hard to see. Maybe shift the range from $r=0.2$.}
		}
    \label{results} 
    \end{center}
\end{figure}

Fig.~\ref{results} shows distribution functions for structural and
energetic properties. We first analyze the hold-out validation data at $T=568$\,K (right
column), the temperature at which DeepBackmap was trained on.
Our method generates configurations that are
remarkably close to the reference Boltzmann distribution (``AA''),
especially when considering the current state of the art.
The distributions of intramolecular carbon backbone angle and dihedral show very good
agreement (Fig.~\ref{results}a--d). On the other hand, the baseline
method displays too narrow distributions and spurious peaks. While the distribution for the carbon
improper dihedral of the aromatic structure is slightly too narrow, we
emphasize the small range of angles (Fig.~\ref{results}e--f), due to
the imposed planarity of the ring. The baseline method significantly
suppresses fluctuations around the planar structure.

The Lennard-Jones energies shown in Fig.~\ref{results} (g--h) obtained
for each chain separately also match remarkably well with the
reference distribution---this aspect is of tremendous importance to
generate well-equilibrated structures in a condensed environment. We
do observe slightly large high-energy tails, often due to an
accumulation of errors of misplaced atoms impacting the subsequent
placement of neighbors in our autoregressive approach. On the other
hand, the baseline model systematically and drastically
over-stabilizes the system. This results from the energy-minimization
scheme, which fails to prepare the structure for a specific canonical
state point. For this reason, state-of-the-art backmapping schemes
require extensive MD simulations, including lengthy heating procedures
and thermostat/barostat equilibration, before offering a starting
point for a production run.

\subsection{Transferability to low temperatures}

While we fix the original training of DeepBackmap to the
high-temperature ensemble, we hereby test it at low temperature
($T=313$\,K), \emph{without} reparametrization. Beyond a mere shift in
temperature, the system undergoes a phase transition, going from an
amorphous phase to a crystalline state with different polymorphs. The
distributions in Fig.~\ref{results}a--g (left column) show remarkably
accuracy: DeepBackmap retains its performance displayed for the
training temperature. Upon cooling the distributions do show a number
of significant changes: narrower distribution in the angle, vanishing
of the side peak in the backbone dihedral, and large shift of the
Lennard-Jones energies.

The transferability of DeepBackmap is highlighted when compared to the
baseline model, which retains much of its features found at high
temperature. This is especially apparent for the side peak of the
backbone dihedral.

\subsection{MD simulation}

Backmapped structures are often used as starting points for further MD
simulations. State-of-the-art backmapping schemes rely on lengthy
preparations to offer a starting point for a production run, such as a
heat-up phase and thermostat/barostat equilibration. 

Fig.~SI.7 displays the evolution of the potential energy
during MD simulations \emph{without} heat-up at $T=313$~K starting from structures generated
with the different methods. Initial velocities are generated according to a Maxwell distribution. The evolution of the potential energy of
structures generated with DeepBackmap follow closely the evolution of
reference AA structures. The potential energy reaches a steady value
after $100$~ps. On the other hand, energy minimized structures from
the baseline method settle at significantly higher energies indicating
badly initialized structures that get trapped into local minima with
high energy barriers.

%\begin{figure}[htbp]
%    \center
%    \includegraphics[width=0.9\linewidth]{./images/equilibration/pot}
%    \caption{Potential energy during MD simulations at $T=313$~K
%    starting from backmapped and reference structures without
%    heat-up.}
%    \label{md_pot_t313}
%\end{figure} 

\subsection{Large-scale structural features}

To further evaluate the large-scale structural features, we turn to
the pair correlation function, $g(r)$. Fig.~\ref{results}i--j focuses here on
non-bonded carbon pairs. We can see an excellent agreement between
the reference AA $g(r)$ and the DeepBackmap results for \emph{both}
temperatures. This clearly indicates that the local packing of the
polystyrene chains is well reproduced, even for different state points
that were not used during training. As expected the baseline method
does not reproduce the pair correlation satisfyingly, especially fails
in the crystalline phase.

Beyond the pair statistics, we wish to probe the accuracy of the
reconstruction at higher order. We build a two-dimensional map
representing proximity relationships between condensed-phase
structures. We focus on the local environment around each backbone
carbon atom that directly links to a side chain (i.e., every other
backbone carbon). The pairwise distance between two such environments
is encoded using a similarity kernel based on SOAP
representations.\cite{bartok2013representing} Hydrogens are ignored
from the representation. To compare two structures we compute a
covariance matrix containing all the pairwise distances between atomic
environments, followed by a regularized entropy match
kernel.\cite{de2016comparing} We further apply Sketchmap to obtain a
reduced-dimensional projection of the conformational
space.\cite{tribello2012using, ceriotti2011simplifying}

\begin{figure}[htbp]
    \center
    \includegraphics[width=0.8\linewidth]{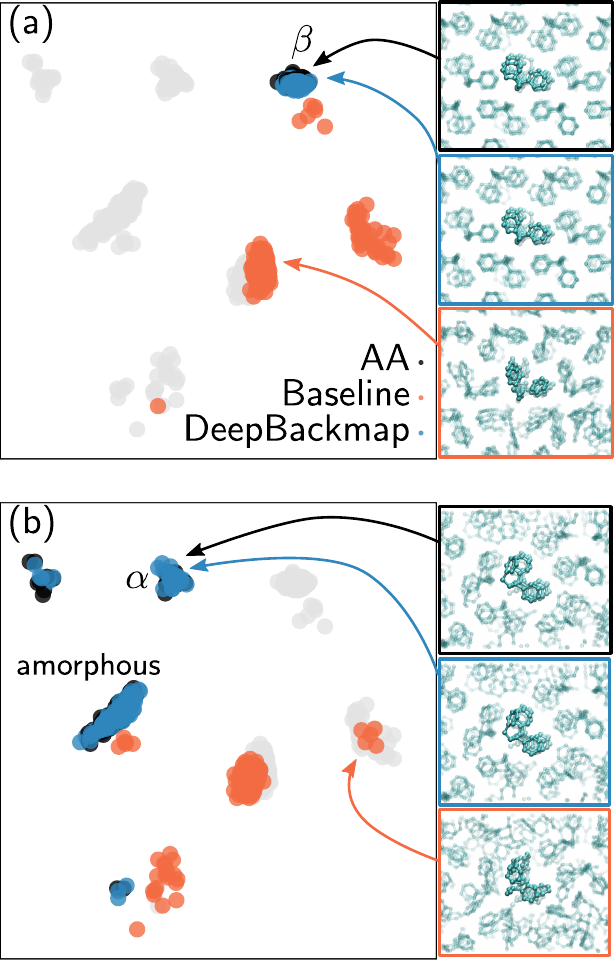}
    \caption{Low-dimensional structural space of condensed-phase
    configurations at (a) $T=313$~K and (b) $T=568$~K. For each panel,
    snapshots are backmapped from identical coarse-grained
    configurations, highlighting the overlap between reference and
    DeepBackmap, but disconnect from the baseline method.} 
    \label{sm} 
\end{figure}

Fig.~\ref{sm} displays a number of clusters that correspond to
different environments. The low-temperature reference data shows a
single cluster (Fig.~\ref{sm}a in black), corresponding to the $\beta$
phase, while the high-temperature reference shows more diversity
(i.e., $\alpha$, amorphous phase, and others). DeepBackmap overlaps
significantly with the reference points at both temperatures,
highlighting the high structural fidelity. This is not the case for
the energy-minimized structures of the baseline model, as they cover
different areas of the low-dimensional map. The baseline also fails to
reproduce the correct number of clusters: at both temperatures the
baseline model displays three to four clusters, highlighting a lack of
temperature sensitivity.

\section{Conclusions}

In this study we propose a new backmapping scheme based on deep neural
networks. The model inserts atomistic details based on large-scale
structures from a coarse-grained snapshot. To this end we use
a conditional generative network where the coarse-grained information
is used as an auxiliary input. We train our model, DeepBackmap,
combining an adversarial loss function with a physical prior. The
method is scalable to arbitrary system sizes since only local
information is used. Our method is able to generate well-equilibrated
high-resolution structures of condensed-phase systems. Critically, and
unlike current methods, our approach does not need molecular dynamics
(MD) simulations to yield the correct Boltzmann distribution. 

We applied our methodology to a complex condensed-phase system made of
syndiotactic-polystyrene chains. The model displays remarkable
transferability properties: while trained solely on high-temperature
melt configurations, DeepBackmap performs well at significantly low
temperatures, where the system is in a crystalline state. This
indicates that the local correlations learned by the model are
transferable across different state points, aided by the physics we
incorporated into the GAN.

We rationalize these remarkable features in terms of scale separation:
the large-scale features are encoded in the coarse-grained
configurations, while the model only need to generate equilibrated
\emph{local} correlations. Local features are less affected by
temperature, since the underlying covalent interactions operate
primarily on an energy scale significantly larger than
$k_\mathrm{B}T$. As such the backmapping operates on two different
sources of information: ($i$) the conditional coarse-grained
configurations and ($ii$) the learned local correlations. Most of the
temperature dependence is carried by the former, such that DeepBackmap
can accurately produce an accurate Boltzmann distribution across a
phase transition from training at a single temperature.

Beyond the evident advantages of generating equilibrated molecular
structures, our approach offers the perspective of a tighter
integration of multiscale models: The information of the
coarse-grained is efficiently recycled into the higher resolution.
Avoiding unnecessary equilibrations upon upscaling will help connect
models at different scales---an important task at the dawn of the
exascale computing era.

\section*{Acknowledgments}

The authors thank Kiran H.~Kanekal, Yasemin Bozkurt Varolg\"une\c{s},
and Arghya Dutta for critical reading of the manuscript. We are
grateful to Chan Liu for providing coarse-grained and atomistic
simulations of syndiotactic polystyrene. This work was supported in
part by the TRR 146 Collaborative Research Center of the Deutsche
Forschungsgemeinschaft as well as the Max Planck Graduate Center. TB
acknowledges financial support by the Emmy Noether program of the
Deutsche Forschungsgemeinschaft (DFG). Part of this research was
performed while MS and TB were visiting the Institute for Pure and
Applied Mathematics (IPAM), which is supported by the National Science
Foundation (Grant No. DMS-1440415).

\section*{Supporting Information}
Link to repository containing trajectories; Fig.~S1: Markov random
field and Gibbs sampling; Fig.~S2: Cutoff and orientation of the
generation; Fig.~S3: Various canonical distributions at $T=453$~K;
Fig.~S4: Radial distribution function at $T=453$~K; Fig.~S5: Sketchmap
at $T=453$~K; Fig.~S6: Convolutional neural network architecture;
Fig.~S7: Potential energy as a function of time starting from
different backmapped structures.

\section*{References}
\bibliography{references}

% \newpage
% \begin{figure*}
%  \includegraphics[width=1.0\textwidth]{./images/architecture.pdf}
% \caption{\label{architecture} }
%  \end{figure*} 
% \newpage

\end{document}